\documentclass[aps,prl,reprint,superscriptaddress,floatfix, citeautoscript]{revtex4-1}
\pdfoutput=1
\usepackage{amsmath}
\usepackage{amssymb}
\usepackage{graphicx}

\newcommand{\beq}{\begin{equation}}
\newcommand{\eeq}{\end{equation}}
\newcommand{\beqs}{\begin{equation*}}
\newcommand{\eeqs}{\end{equation*}}
\newcommand{\beqy}{\begin{eqnarray}}
\newcommand{\eeqy}{\end{eqnarray}}
\newcommand{\bit}{\begin{itemize}}
\newcommand{\eit}{\end{itemize}}

\begin{document}

\title{Dynamic multiferroicity of a ferroelectric quantum critical point}

\author{K. Dunnett}
\affiliation{Nordita, KTH Royal Institute of Technology and Stockholm University, Roslagstullsbacken 23, SE-106 91 Stockholm, Sweden}
\author{{J.-X. Zhu}}
\affiliation{T-4 and CINT, Los Alamos National Laboratory, Los Alamos, New Mexico 87545, USA}
\author{N. A. Spaldin}
\affiliation{Materials Theory, ETH Zurich, Wolfgang-Pauli-Strasse 27, CH-8093 Z\"{u}rich, Switzerland}
\author{V. Juri{\v c}i{\'c}}
\affiliation{Nordita, KTH Royal Institute of Technology and Stockholm University, Roslagstullsbacken 23, SE-106 91 Stockholm, Sweden}
\author{A. V. Balatsky}
\affiliation{Nordita, KTH Royal Institute of Technology and Stockholm University, Roslagstullsbacken 23, SE-106 91 Stockholm, Sweden}
\affiliation{Department of Physics, University of Connecticut, Storrs, CT 06269, USA}

\date{\today} 

\begin{abstract}
Quantum matter hosts a large variety of phases, some coexisting, some competing; when two or more orders occur together, they are often entangled and cannot be separated. Dynamical multiferroicity, where fluctuations of electric dipoles lead to magnetisation, is an example where the two orders are impossible to disentangle. Here we demonstrate elevated magnetic response of a ferroelectric near the ferroelectric quantum critical point (FE QCP) since magnetic fluctuations are entangled with ferroelectric fluctuations. We thus suggest that any ferroelectric quantum critical point is an \textit{inherent} multiferroic quantum critical point. We calculate the magnetic susceptibility near the FE QCP and find a region with enhanced magnetic signatures near the FE QCP, and controlled by the tuning parameter of the ferroelectric phase. The effect is small but observable - we propose quantum paraelectric strontium titanate as a candidate material where the magnitude of the induced magnetic moments can be $\sim 5 \times 10^{-7} \mu_{B}$ per unit cell near the FE QCP. 
\end{abstract}

\maketitle

Quantum matter exhibits a plethora of novel phases and effects upon driving~\cite{NatMat.16.1077}, one of which is the strong connection between the quantum critical point (QCP) of one order parameter and the presence of another phase. The discussion has often focussed on the relation between superconductivity and one or more magnetic phases~\cite{NatPhys.4.186, AnnRevCondMatPhys.5.113, RepProgPhys.79.082501}. However, other fluctuation-driven phase transitions, for example nematic phases in iron-based superconductors~\cite{NatPhys.10.97, AnnRevCondMatPhys.5.113}, have also received significant attention. We focus here on the ferroelectric (FE) QCP which is a key part of the discussion of FE behaviour, particularly in displacive quantum paraelectrics~\cite{NatPhys.10.367, RepProgPhys.80.112502}. The behaviours that may occur near or as a result of such an FE QCP have been explored in various contexts ~\cite{NatPhys.10.367, RepProgPhys.80.112502, PhysRevB.79.075101, PhysRevLett.115.247002, JPhysCondMat.27.395901, NatPhys.13.7.643, NMat.31.12.2018, PhysRevMaterials.2.104804}, and the list of systems where the effects of quantum fluctuations can be observed is expanding, with temperatures up to $\sim 60 \mathrm{K}$ in some organic charge-transfer complexes~\cite{NatComm.6.7469, JPhysCondMat.27.395901}. 

The concept of dynamical multiferroicity was introduced recently as the dynamical counterpart of the Dzyaloshinskii-Moriya mechanism, reflecting the symmetry between electric and magnetic properties~\cite{PhysRevMat.1.014401}. In the Dzyaloshinskii-Moriya mechanism~\cite{JPhysChemSol.4.241, PhysRevLett.4.228, PhysRev.120.91}, ferroelectric polarisation is caused by a spatially varying magnetic structure, leading to strong coupling between ferroelectricity and magnetism~\cite{PhysRevLett.95.057205, PhysRevB.73.094434, NatMat.6.13}. In the related phenomenon of dynamical multiferroicity, magnetic moments $\mathbf{m}$ can be induced by time-dependent oscillations of electric dipole moments $\mathbf{p}$:
\beq
\mathbf{m} = \lambda\, \mathbf{p} \times \partial_t\mathbf{p} = C\, \mathbf{n} \times \partial_t \mathbf{n}.
\label{eq:DefMfromPt}
\eeq
For magnetic moments to be induced, $\mathbf{p}$ has to exhibit transverse fluctuations; we therefore focus on rotational degrees of freedom of electric dipole moments \cite{PhilMag.89.2079}. The unit direction vector of the constant amplitude electric dipole moment is $\mathbf{n} \equiv \mathbf{n}(\mathbf{r},t)$, with time derivative $\partial_t\mathbf{n}$, and $C=\lambda {|\mathbf{p}|}^2$ in terms of the electric dipole moments $\mathbf{p}$ (we use estimates from uniform polarisation $P_0 = |\mathbf{p}|V$ with volume $V$ in FE phases), and coupling $\lambda = \pi/e$, with $e$ the electric charge. Generally, we expect that orders entangled with the underlying static order can be excited dynamically. One possibility is to use external driving mechanisms such as light, magnetic field or lattice strain to induce transient excitations of the entangled orders \cite{NatPhys.4.186}. The present work addresses the complementary case where inherent FE quantum fluctuations induce entangled ferromagnetic order fluctuations without any external drive.

\begin{figure}[ht!]
\includegraphics[width=\columnwidth]{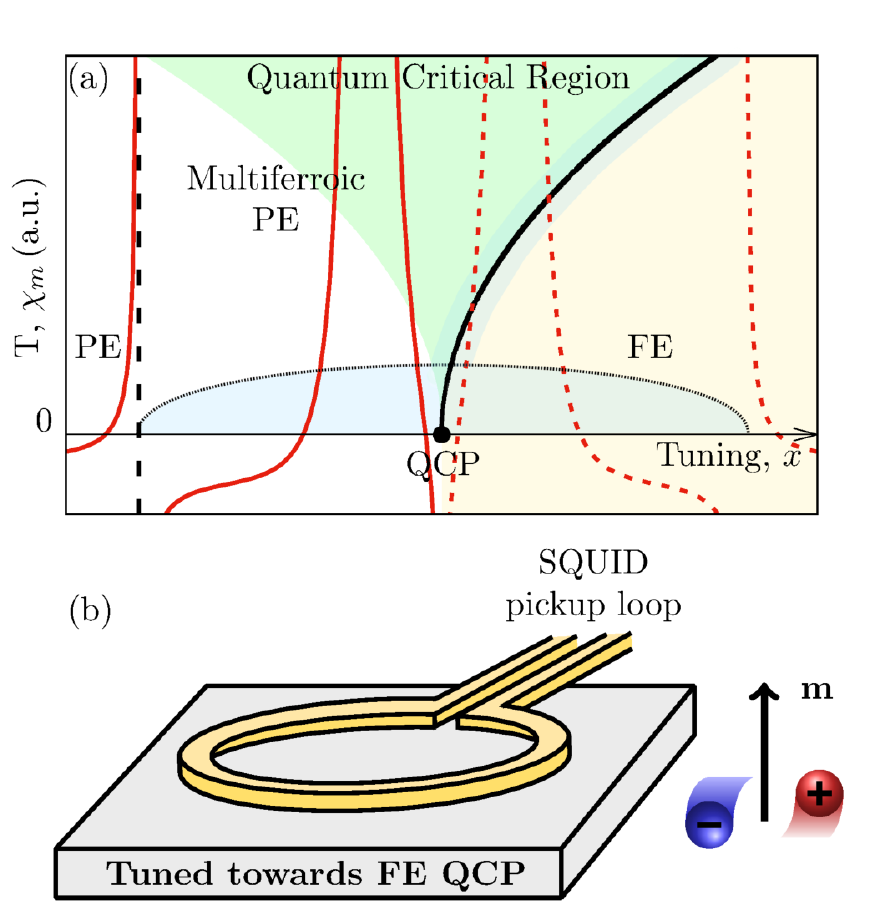}
\caption{(a) Phase diagram near a ferroelectric quantum critical point (QCP, where $x=x_{cr}$) with the magnetic susceptibility (red line, dashed in FE phase) at $\omega = 0.5\omega_0$ which diverges at the vertical dashed line, leading to a new `Multiferroic PE' phase. The ferroelectric quantum critical region (pale green) is now a dynamical multiferroic quantum critical region. In both the PE (white background) and FE (yellow background) phases, qualitatively similar behaviours of $\chi_m$ are expected, despite the different underlying orders. The blue shading indicates the main regions where induced magnetic signatures are expected: a dome around the FE QCP due to quantum fluctuations; a narrow band around the finite temperature phase transition line due to thermally induced fluctuations. (b) A simple experiment using a SQUID could detect magnetic signatures resulting from rotating electric dipoles in a system towards its ferroelectric phase transition. Here, the electric dipoles are constrained to the horizontal plane and lead to an out-of-plane magnetic moment $\mathbf{m}$ and susceptibility. 
\label{fig:PhaseDiagram}}
\end{figure}

\begin{figure}[ht!]
\includegraphics[width=\columnwidth]{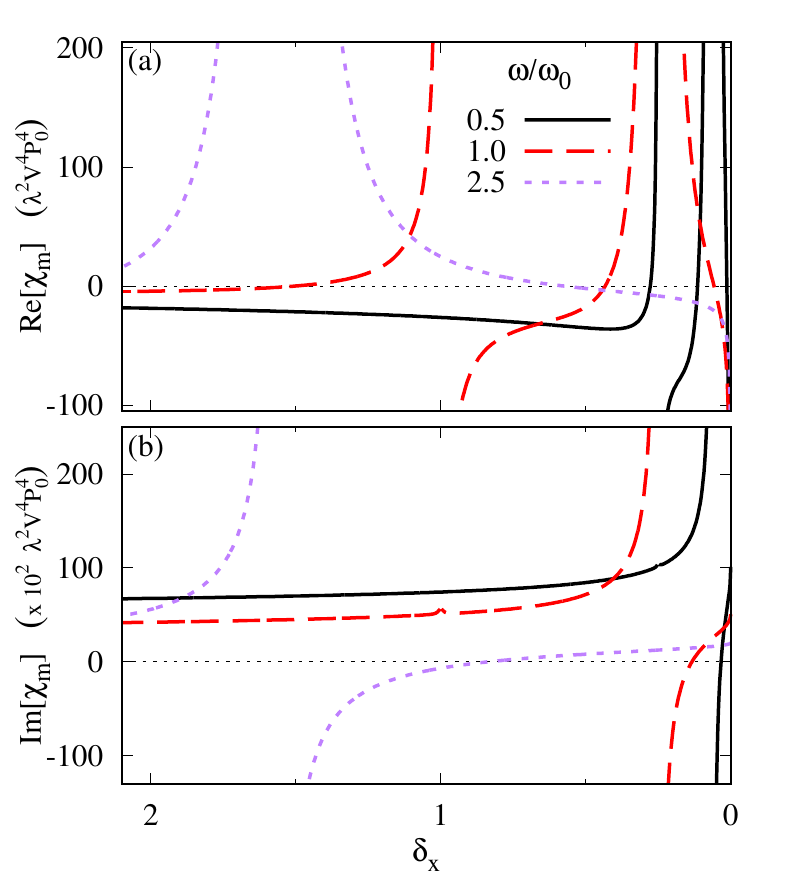}
\caption{The total magnetic susceptibility in units of the common prefactor $C^2 = \lambda^2 V^4 P_0^4$, and with $\Lambda^3/\omega_0 = 1 \times 10^5$, as a function of $\delta_x$ at several energies. (a) the real part; (b) the imaginary part. The behaviour in the FE phase $\delta_x<0$ is expected to share the main qualitative features despite the underlying order. The effects of changing the $\chi^{(2)}$ prefactor $\Lambda^3/\omega_0$ and the individual contributions of $\chi^{(1)}$ and $\chi^{(2)}$ are discussed in the SM, \S IV~\cite{SuppMat}. \label{fig:ChiFunctionsWQCP}}
\end{figure}

\begin{figure}[t!]
\includegraphics[width=\columnwidth]{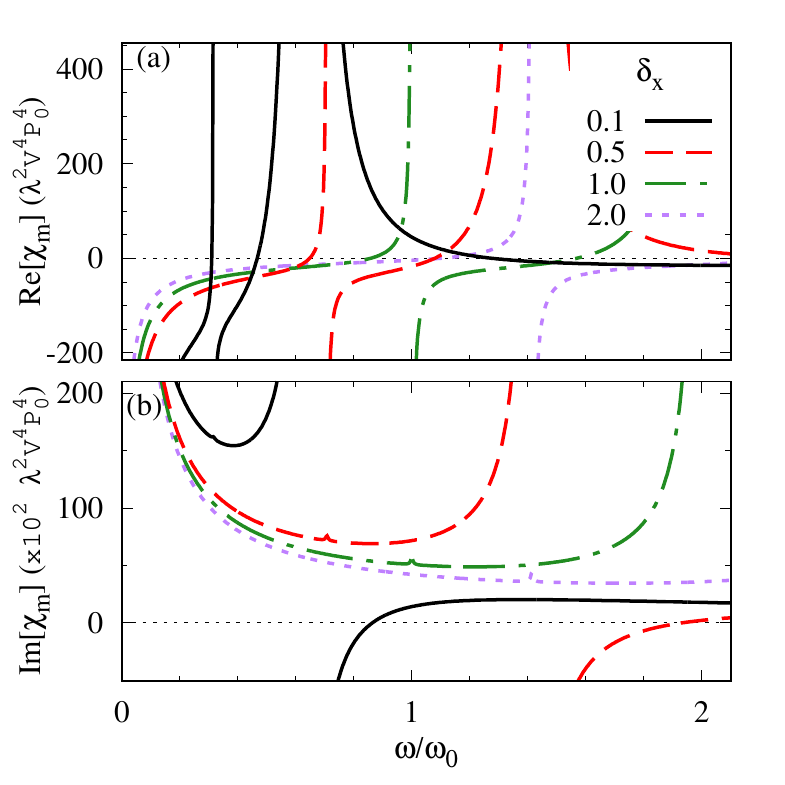}
\caption{Total magnetic susceptibility, in units of $C^2 =\lambda^2 V^4 P_0^4$, and with $\Lambda^3/\omega_0 =1\times 10^5$, as a function of $\omega/\omega_0$ at several distances from the FE QCP. (a) real part; (b) imaginary part. The peaks in the imaginary part ($\delta-$functions plotted as Lorentzian functions) corresponding to the divergence and sign change of the real part are weak for $\Lambda^3/\omega_0 = 1\times 10^5$ used; their presence is more easily seen at smaller values of $\Lambda^3/\omega_0$ (SM, \S IV \cite{SuppMat}). \label{fig:ChiFunctionsWEnergy}}
\end{figure}

In this Letter, we demonstrate that i) the fluctuating dipoles can induce magnetic fluctuations that surround the FE QCP, as shown in Fig.~\ref{fig:PhaseDiagram}a. The mechanism for this effect is the induction of magnetic moments by fluctuating  electric dipoles, described by Eq.~\eqref{eq:DefMfromPt}, near the FE QCP and therefore describes \emph{inherent} dynamic multiferroicity. We support this scenario by calculating the magnetic susceptibility that, as we show, diverges in the paraelectric phase (Figs.~\ref{fig:ChiFunctionsWQCP} and~\ref{fig:ChiFunctionsWEnergy}), indicating a transition to a new regime, labelled `Multiferroic PE' in \mbox{Fig. \ref{fig:PhaseDiagram}a}. We thus surmise that any FE QCP is a multicritical multiferroic (MF) QCP with elevated magnetic fluctuations. We stress that the proposed effect is not due to permanent intrinsic magnetic moments, for example from unpaired electrons on ions, but arises solely due to dynamics of the ferroelectric order. While the proposed effect is general, we will consider the specific implications for magnetism in strontium titanate (STO) and provide estimates relevant to STO. ii) Within the approximations used, the application of a magnetic field $\mathbf{B}$ does not introduce a static, $\mathbf{B}$-dependent mass term to the effective action for $\mathbf{p}$, and the position of the FE QCP is therefore independent of $\mathbf{B}$. The Zeeman splitting  of the FE active phonon modes \cite{PhilMag.89.2079, PhysRevMat.1.014401} meanwhile does affect the magnetic susceptibility $\chi_m$ and, in higher order approximations, is expected to lead to a $\mathbf{B}^2$ term in the free energy, shifting the FE QCP. iii) We estimate the typical induced magnetic moment from a single rotating electric dipole to be $|\mathbf{m}| \approx 5 \times 10^{-7} \mu_B$, where $\mu_B$ is the Bohr magneton. This is for coupling $\lambda = \pi/e$ and a dipole with charge $4e$ and length $1\times 10^{-2}${\AA}, rotating with frequency $0.5 \,\mathrm{THz}$, typical of the titanium displacements \cite{SolStatComm.9.191, PhysRevLett.82.3540, PhysRevB.64.174104, PhysRevB.95.054107} and the ferroelectric phonon modes in STO~\cite{JPhysSocJpn.26.396, PhysRevB.13.271, EPL.50.688} (supplemental material (SM), \S I ~\cite{SuppMat}). The overall contribution of the fluctuating FE order is diamagnetism near the FE QCP.

\textit{Model:} The system considered consists of fluctuating electric dipoles close to the PE-FE phase transition, inducing a magnetic moment via Eq. \eqref{eq:DefMfromPt}. In the absence of external fields, the generic description of the system of rotating electric dipoles consists of the paraelectric phase: $L_{PE} = \left(\omega^2 - \omega_q^2 \right)\mathbf{p}_{\omega, q} \mathbf{p}_{-\omega, -q} $, where $\omega_q$ is the dispersion of the phonon mode relevant for ferroelectricity, $\mathbf{p}_{\omega, q}$ is the rotating electric dipole moment written in Fourier (energy $\omega$ - momentum $q$) space. The paraelectric phase has negligible intrinsic magnetic contribution and we therefore ignore intrinsic magnetisation altogether. However, the dynamic induction of $\mathbf{m}$, Eq. \eqref{eq:DefMfromPt}, will lead to magnetic susceptibility of the paraelectric near the FE QCP.

The interaction between induced magnetic moments can be neglected in the PE phase since the lowest order contribution ${|\mathbf{m}|}^2 \propto {|\mathbf{p}|}^4$. We assume optical phonons, relevant for the PE-FE transition in STO~\cite{JPhysSocJpn.26.396}, with dispersion $\omega_q$ given by:
\beq
\omega_q^2 = \omega_0^2 \left(1-\frac{x}{x_{cr}}\right) + b q^2 = \omega_0^2 \delta_x + b q^2
\label{eq:OptPhonDisp},
\eeq
where $\delta_x = x/x_{cr}$ describes the distance to the \textit{ferroelectric} QCP at $x_{cr}$; $\omega_0$ is energy at the zero momentum of the soft mode when $x=0$, i.e. with no driving of the system towards the FE QCP, and $q$ is the momentum. $x$ is a tuning parameter that controls the paraelectric-ferroelectric phase transition at zero temperature, such as doping. If the system is very close to the FE QCP, the momentum dependence is negligible and a flat dispersion with $b = 0$ can be used. The system is paraelectric for $\delta_x>0$ and ferroelectric when $\delta_x<0$.

Although in reality both amplitude and directional fluctuations of $\mathbf{p}$ are present near the FE QCP, we will ignore the amplitude fluctuations, so the time dependence of $\mathbf{p}$ is contained entirely in the unit direction vector $\mathbf{n}$. In this model, at the boundary between the PE and FE phases instead of $|\mathbf{p}|\rightarrow 0$, the dipoles rotate. That is: in the PE phase, finite-sized electric dipoles are present, but not aligned so the net polarisation is zero, and in the FE phase the dipoles align. $\mathbf{n}$ is linearised as: $\mathbf{n} = \mathbf{n}_0 + \mathbf{\tilde{n}}(t)$ with $\partial_t \mathbf{n}_0 = 0$ and $\langle \tilde{\mathbf{n}} \rangle = 0$. The zero-temperature Green's function of the $\mathbf{\mathbf{n}}$ field in $\omega-q$ space reads
\beq
\langle \tilde{n}_{\omega}^j \tilde{n}_{-\omega}^m \rangle = A_{j}\delta_{jm} G(i\omega,q), \label{eq:GreenFunctionDef}
\eeq
with $A_j$ as a constant factor. To find dynamic susceptibilities, we use the retarded Green's function $G^R$, obtained by analytical continuation to real frequencies ($i\omega\rightarrow\omega+i\eta$, SM, \S II~\cite{SuppMat}):
\beq
G^R(\omega,q) = \mathrm{Re}\left(\frac{1}{\omega_q^2 - \omega^2}\right) + \frac{i\pi}{2\omega_q}\left[\delta(\omega_{q} -\omega) - \delta(\omega_{q} + \omega) \right]. \label{eq:GreenFunctionUsed} \qquad
\eeq
We now calculate the magnetic susceptibility $\chi_m$ in the PE phase: 
\beq
\chi_m = \langle \mathbf{m}(r_1, t_1) \mathbf{m}(r_2, t_2) \rangle
\equiv \chi^{(1)} + \chi^{(2)} \label{eq:MagSuscDef}
\eeq
where $\mathbf{m}$ is given by Eq. \eqref{eq:DefMfromPt}. The two contributions to $\chi_m$ are $\chi^{(1)} \propto \langle \tilde{n}^k \tilde{n}^n \rangle$ and $\chi^{(2)} \propto \langle \tilde{n}^j \tilde{n}^k \tilde{n}^m \tilde{n}^n \rangle$.

The quadratic contribution in $\omega-q$ space is:
\beq
\chi^{(1)}_{il} = C^2 n_0^j n_0^m A_k \epsilon_{ijk}\epsilon_{lmk}\omega^2 G^R(\omega),
\eeq
with $G^R(\omega)$ given by Eq. \eqref{eq:GreenFunctionUsed}, the factor $C^2=\lambda^2V^4P_0^4$ gives the size of the magnetic susceptibility in terms of the coupling $\lambda$ for the induced magnetic moments, and the polarisation $P_0$ of a sample of volume $V$. $n_0^i$ are the components of $\mathbf{n}_0$ around which the fluctuations are expanded. The factor $\omega^2$ comes from the Fourier transform of $\langle \partial_t \tilde{n}^k \partial_t \tilde{n}^n \rangle$.

The quartic contribution to the magnetic susceptibility corresponds to a one-loop diagram as discussed in the SM, \S III~\cite{SuppMat}, with the real part given by
\beq
\mathrm{Re}[\chi^{(2)}_{ii}] = - \frac{C^2 \delta_{il} A_j A_k\Lambda^3}{8\pi \omega_x} f(\omega)\label{eq:ReChi2}
\eeq
where $f(\omega)$, given in full in the SM, \S III~\cite{SuppMat}, contains $\delta$-functions at $2\omega_0\sqrt{\delta_x}\pm \omega $ and $\omega$ with weights $\omega$ or $\omega_0\sqrt{\delta_x}$, and $\Lambda$ is a momentum cut off. The imaginary part is:
\beq
\mathrm{Im}[\chi^{(2)}_{ii}] =\frac{C^2\delta_{il} A_j A_k\Lambda^3}{\pi^2(\omega^2 - 4\omega_x^2)} \left\{\frac{\omega^2-2\omega_x^2}{2\omega} - \omega_x \right\}. \label{eq:ImChi2}
\eeq
If the energy $\omega$ is written in terms of the $q=0$ phonon energy $\omega_0$, the size of the $\chi^{(2)}$ contribution is determined by $\Lambda^3/\omega_0$. In STO, areas of coherent fluctuations are limited to tetragonal domains, $\sim 10 \mu\mathrm{m} $~\cite{PhysRevB.94.174516}, in which case $\Lambda^3/\omega_0 \sim 5\times 10^5$, for $\omega_0 = 0.5\mathrm{THz}$, as suitable for the ferroelectric optical phonons in STO. Further, the distribution and size of tetragonal domains can be controlled by both applied electric fields~\cite{JPhysCondMat.22.235903, NatMat.12.1112} and pressure~\cite{JPhysCondMat.22.235903}.

\textit{Results:} The total magnetic susceptibility $\chi_m$ from Eq. \eqref{eq:MagSuscDef} is plotted in Figs. \ref{fig:ChiFunctionsWQCP} and \ref{fig:ChiFunctionsWEnergy} with the overall scale given by the shared prefactor $C^2=\lambda^2V^4P_0^4$ set to unity in all plots. In STO samples, the value of $C^2$ can be estimated from experimental data of samples tuned through the FE phase transition by applied strain or $^{18}\mathrm{O}$ isotope substitution, which indicates the possible size of the dipole moments in the PE phase: $C^2 \sim 2\times 10^{-3} \mathrm{C^2m^4}$, for bulk STO crystals, and $C^2\sim 4 \times 10^{-34} \mathrm{C^2m^4}$ for $10\mu\mathrm{m}$ tetragonal domains ~\cite{SolStatComm.9.191, PhysRevLett.82.3540, PhysRevB.64.174104}.Considering a sample with a single induced magnetic moment of $5\times 10^{-7} \mu_B$ per unit cell and a sample volume of $1\mu m^3$, smaller than the tetragonal domains in STO, gives a sample magnetic moment of $8\times 10^4 \mu_B$, well within spin sensitivities of $200\mu_B/\sqrt{\mathrm{Hz}}$ of current SQUIDs \cite{RevSciInstrum.79.053704}.

We consider tuning towards the FE QCP at a constant energy (fixed $\omega/\omega_0$) first. In Fig. \ref{fig:ChiFunctionsWQCP}a, far from the FE QCP, the system is dielectric with $\mathrm{Re}[\chi_m]>0$ but not large. On moving towards the FE QCP, $\chi_m$ diverges and changes sign at $\delta_x = (\omega/\omega_0)^2$; this indicates a phase transition into a region where magnetic signatures can be expected. As the energy is decreased, the divergence moves towards the FE QCP and the magnetic features are confined into a narrower range of the tuning parameter. 

There are two contributions to the peaks in the real part of the susceptibility: one is from the poles in $\mathrm{Re}[\chi^{(1)}]$ resulting in the large derivative feature at $\delta_x= (\omega/\omega_0)^2$; the other comes from the $\delta$-functions in $\mathrm{Im}[G^R]$ that lead to peaks in $\mathrm{Re}[\chi^{(2)}]$ at $\delta_x = (\omega/2\omega_0)^2$. After the initial divergence, $\mathrm{Re}[\chi_m]$ is negative apart from the sharp peak at $\delta_x = (\omega/2\omega_0)^2$, below which it quickly reaches a constant negative value independent of $\omega$.

The imaginary part of $\chi_m$, plotted in Fig. \ref{fig:ChiFunctionsWQCP}b, also diverges as expected at the border of the magnetic region. This is followed by a divergence at $\delta_x = (\omega/2\omega_0)^2$ corresponding to the peaks originating from $\chi^{(2)}$ in the real part. In the limit of $\delta_x \rightarrow 0$, $Im[\chi_m]$ reaches a positive value that depends on the energy $\omega$ considered. It is important to note that the details of both the real and imaginary parts once the magnetic phase transition has been passed, that is for $\delta<(\omega/\omega_0)^2$, are determined solely by the higher order $\chi^{(2)}$ contributions.

Changing energy while at a fixed distance from the FE QCP is considered in Fig. \ref{fig:ChiFunctionsWEnergy}. The peaks and divergences at finite $\omega$ are exactly those seen in Fig. \ref{fig:ChiFunctionsWQCP}, with an extra, artificial, divergence of both the real and imaginary parts at $\omega=0$, originating from calculating $\chi^{(2)}$ in the continuum limit. At the lowest energies, $Re[\chi_m]<0$, it then increases and diverges at $\omega/\omega_0 = \sqrt{\delta_x}$, thus signalling the phase transition with magnetic signatures expected above a critical energy scale. We note that upon increasing $\delta_x$ to move away from the QCP, the onset of the transition moves to higher energy. The energy dependence of the size of the imaginary part is seen particularly clearly in Fig. \ref{fig:ChiFunctionsWEnergy}b. In the FE phase, we expect qualitatively similar features, despite the underlying FE order, due to fluctuations of the ordered dipoles.

A magnetic field $\mathbf{B}$, applied perpendicular to the plane of the rotating dipoles, will have two effects. Firstly, the phonon Zeeman effect splits the phonon modes with a linear dependence on $\mathbf{B}$ \cite{PhysRevMat.1.014401} and moves the divergence of $\chi_m$ [which occurs at $\delta = (\omega/\omega_0)^2$ for $\mathbf{B} = 0$]. Secondly, an additional term in the Lagrangian for the interaction of the induced magnetic moments with the $\mathbf{B}$, $\mathbf{B}\cdot\mathbf{m} = \lambda \mathbf{B}\cdot (\mathbf{p} \times \partial_t \mathbf{p})$ \cite{PhilMag.89.2079}, can be treated as a perturbation to the paraelectric system. Calculating the corresponding second order diagram (SM, \S V \cite{SuppMat}) does not introduce a static, $\mathbf{B}$-dependent mass term, but may do so at higher orders.

\textit{Experimental proposal:} Strontium titanate (STO) may be a suitable candidate material for the observation of magnetic signatures on tuning towards the FE QCP because of its incipient ferroelectric nature below $\sim 35$ K and its quantum paraelectric nature below $4$ K~\cite{PhysRevB.19.3593} where the zero-point motion of the soft transverse optical phonon mode is high enough to prevent ferroelectricity even at zero temperature~\cite{PhysicaB.219.577}. In $^{18}\mathrm{O}$ substituted STO, $\omega_{q=0}(T)$ becomes constant below $4-10\mathrm{K}$ depending on the distance from the FE QCP ~\cite{PhysRevB.51.8046, PhysRevB.69.024103, PhysRevLett.96.227602, PhysRevLett.99.017602, PhysScripta.89.025702}. Thus, rotating electric dipoles could be present over an appreciable temperature range. Additional flexibility exists because several methods are available for tuning STO towards the FE QCP, such as Ca doping~\cite{PhysRevLett.52.2289}, $^{18}$O substitution~\cite{PhysRevLett.82.3540, PhysRevB.64.174104, ApplPhysLett.76.221}, strain or applied pressure~\cite{PhysRevB.13.271, SolStatComm.9.191}.

A simple experimental set up, consisting of a superconducting quantum interference device (SQUID) above an STO sample, that may permit the observation of the region of pronounced magnetic fluctuations is sketched in Fig. \ref{fig:PhaseDiagram}b. Strain is a particularly flexible means of tuning STO samples towards the FE QCP, and biaxial strain in STO thin films can confine polarisation to the plane perpendicular to the tetragonal c-axis, but does not unambiguously determine the polarisation direction~\cite{PhysRevB.61.825, PhysicaB.406.4145, JMatSci.49.5978, PhysRevB.67.014105}, a favourable condition for the observation of the magnetic signatures proposed here. Although strained STO is considered here, other FE QCPs and tuning mechanisms could be studied, e.g.: $\mathrm{Ca}_{1-x} \mathrm{Pb}_x \mathrm{TiO}_3 $ ~\cite{ApplPhysLett.81.886}, strained $\mathrm{KTaO}_3$~\cite {PhysRevLett.104.227601}. The quantum dipole phase of the triangular lattice Mott insulator $\kappa-\mathrm{(BEDT-TTF)}_2\mathrm{Hg(SCN)_2Br}$~\cite{Sci.360.6393.1101} may also exhibit magnetic signatures of inherent dynamical multiferroicity. The crucial ingredient for inherent dynamical multiferroicity is incipient ferroelectricity (or quantum paraelectricity) and only weak anisotropy between at least two in-plane polarisation directions, to allow the fluctuations to well-defined circulating ions.

\textit{Discussion:} 
Including the long range interactions between electric dipoles, such as those resulting from twin boundaries between tetragonal domains with differently oriented c-axes~\cite{PhysRevLett.116.257601, NatMat.16.1203}, would introduce off-diagonal terms to the Green's function~\cite{PhysRevB.67.014105}. The immediate effect is a non-zero average magnetisation $\langle \mathbf{M} \rangle \propto \langle \mathbf{n} \times \partial_t\mathbf{n} \rangle$. Alongside this, the off-diagonal components of the dielectric susceptibility $\chi_e^{ij} = \langle p_i p_j \rangle \propto \langle \tilde{n}^i \tilde{n}^j \rangle$ would also be non-zero at the twin boundaries, leading to a finite Kerr effect~\cite{PhysRevB.96.214423}. Further, the motion of twin boundaries  may be a means to induce relevant fluctuations of the electric dipoles~\cite{PhysRevB.89.184104}. Scanning SQUID measurements able to resolve the individual tetragonal domains would be required to investigate the effects of domain structures on the magnetic signals. Again, STO is a potential candidate material since tetragonal domains form naturally on cooling through the antiferrodistortive structural phase transition at 105 K and their distribution can be controlled by applied pressure~\cite{JPhysCondMat.22.235903}.

The situation examined here is distinct from that recently considered in the context of multiferroic criticality~\cite{NMat.31.12.2018} and other systems where the quantum critical points of two or more types of order can be tuned by the same or different parameters leading to a fan where the quantum fluctuations of both orders are important~\cite{NMat.31.12.2018, PhysRevB.96.245104}. In our model, the magnetic order does not exist independently of the ferroelectric order, leading to an FE quantum critical region that is surrounded by a region of strong magnetic fluctuations. While distinct from the nematic phase transitions seen in iron pnictides~\cite{NatPhys.10.97, PhysRevB.85.024534}, the multiferroic paraelectric region is another realization of competing orders near a QCP. The interaction between the induced magnetic moments and an external magnetic field is expected to mostly affect the nature of the FE phase transition, as discussed for magnetic phase transitions~\cite{PhysRevB.82.165128, PhysRevB.85.165111, PhysRevA.92.013606}.

\textit{Conclusions:} We have expanded the framework of dynamic multiferroicity~\cite{PhysRevMat.1.014401}, and predict strongly enhanced ferromagnetic (FM) susceptibility in a paraelectric material near its FE QCP. The induced magnetic susceptibility diverges at a finite distance from the FE QCP. The predicted effect points to another way for entangled quantum orders to appear. On the approach to the FE QCP, the fluctuations of the entangled (FM) order are enhanced as the static FE order develops quantum fluctuations. We thus suggest that any FE QCP may be an \textit{inherent} multiferroic QCP with entangled ferroelectric and (much weaker but present) ferromagnetic fluctuations. We expect magnetic signatures of fluctuating dipoles to be observable experimentally, e.g. in SQUID measurements and could lead to additional signatures in optical Kerr and Faraday effects. Our results are applicable to any ferroelectric-paraelectric transition including classical transitions at finite temperatures, where the fluctuations will be confined to a narrow Ginzburg-Levanyuk region near the transition. The effect will become pronounced near the $T=0$ QCP. Finally, to illustrate this scenario, we have considered STO as a system that can be tuned towards its FE QCP.

\begin{acknowledgments} 
We are grateful to G. Aeppli, J. Lashley, and I. Sochnikov for useful discussions. The work was supported by the US DOE BES E3B7, by VILLUM FONDEN via the Centre of Excellence for Dirac Materials (Grant No. 11744) and by The Knut and Alice Wallenberg Foundation (2013.0096).
\end{acknowledgments}

%

\clearpage

\newpage
\appendix 
\section*{Supplemental material}

\section{I: Coupling strength}
To calculate the strength of the coupling in the dynamical multiferroic set up, we consider a current $I$ flowing round a loop of radius $r=a_B$ (the Bohr radius) and a period $\tau_B$ such that the energy $\hbar/\tau_B$ is one Rydberg $R_y = \hbar^2/2m_e a_B^2$. The magnetic dipole moment is perpendicular to the plane of the loop and has magnitude~\cite{GriffithsEmag}
\beq
m_{B} = I\pi r^2 = \frac{e\pi a_B^2}{\tau_B} = \frac{e\hbar \pi}{2m_e} = \pi \mu_B, \label{eq:BohrMagMom}
\eeq
where $\mu_B = e\hbar/2m_e$ is the Bohr magneton.

The coupling strength $\lambda$ is obtained by equating the magnitude of the magnetic dipole moment of the current loop ($m_{B}$, above) with that of the dynamical multiferroic set up [Eq. \eqref{eq:DefMfromPt} of the main text] for electric dipole moments of charge $e$ and length $a_B$ rotating with period $\tau_B$:
\beq
m_{dyn} = \frac{\lambda e^2 a_B^2}{\tau_B}.
\eeq
Requiring that these magnetic moments are equal, $m_{B} = m_{dyn}$, gives the coupling strength:
\beq
\lambda = \pi\mu_B\frac{\tau_B}{e^2 a_B^2}= \frac{\pi e a_B^2}{\tau_B}\frac{\tau_B}{e^2 a_B^2} = \frac{\pi}{e}.
\eeq
The ratio of the induced magnetic moment of any rotating electric dipole to the Bohr model [Eq. \eqref{eq:BohrMagMom}] is:
\beq
\frac{m_{FE}}{m_B} = \frac{(n_q)^2(n_d)^2 \tau_{B}}{\tau_{F}}
\eeq
where $\tau_{F}$ is the rotation period of the electric dipole(s), $n_q$ and $n_d$ are the charge and size of the dipoles in units of the electron charge and Bohr radius respectively. 

\section{II: Model}

The analytical continuation from the Matsubara frequencies to real frequencies consists of replacing $i\omega$ in the Green's function by $\omega + i \eta$~\cite{Mahan_ManyParticlePhysics}:
\beqy
G(i\omega, q)  &=& \frac{1}{\omega_q^2 - (i\omega)^2} \rightarrow G^R(\omega, q) \nonumber \\
G^R(\omega, q) &=& \frac{1}{\omega_q^2 -(\omega + i\eta)^2} \nonumber \\ 
&=& \frac{1}{2\omega_q}\left[\frac{1}{\omega_q + \omega + i\eta} + \frac{1}{\omega_q - \omega - i\eta}\right]
\eeqy
This is then evaluated using the principal value integrals to give:
\beqy
G^R(\omega,q) &=& \frac{1}{2\omega_{q}}\Bigg\{\frac{1}{\omega_{q} + \omega} + \frac{1}{\omega_{q} -\omega} \nonumber \\
&& \qquad +\; i \pi\left[\delta(\omega_{q} -\omega) - \delta(\omega_{q} + \omega) \right] \Bigg\} \nonumber \\
&=& \frac{1}{\omega_{q}^2 -\omega^2} + \frac{i \pi}{2\omega_q}\big[\delta(\omega_{q} -\omega) - \delta(\omega_{q} + \omega) \big], \qquad \label{eq:GreensFunctReDefSM}
\eeqy
where $\omega_q$ is the dispersion of the ferroelectric (FE) phonons: $\omega_q = \sqrt{\omega_0^2 \delta_x + b q^2}$. Near the ferroelectric quantum critical point (FE QCP), the momentum dependence can be neglected~\cite{PhysRevB.67.014105}.

\section{III: Calculation of $\chi^{(2)}$}

The second contribution to the magnetic susceptibility is:
\beq
\chi^{(2)}_{m, il} = C^2 \epsilon_{ijk}\epsilon_{lmn}\langle \tilde{n}^j(t_1) \partial_{t_1} \tilde{n}^k(t_1) \tilde{n}^n(t_2) \partial_{t_2}\tilde{n}^m(t_2) \rangle \label{eq:QuarticTermSM}
\eeq
where the temporal (and spatial) arguments have been included explicitly ($(t_1) \Rightarrow (r_1, t_1)$), and $C^2 = \lambda^2 V^4 P_0^4$. The easiest way to evaluate this is to recognise that the angular bracket corresponds to the loop of the diagram:
\begin{center}
\includegraphics[width=0.8\columnwidth]{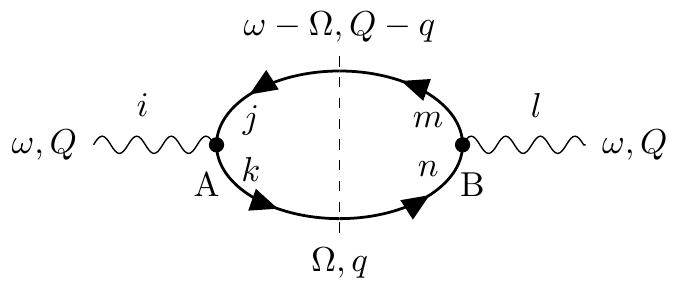}
\end{center}
in which the wiggly lines represent magnetic propagators while the plain curves with direction arrows are ferroelectric propagators. The equivalence of the internal lines means $\delta_{jm}$ and $\delta_{kn}$ and the $\epsilon_{ijk}\epsilon_{lmn}$ prefactor becomes $\epsilon_{ijk}\epsilon_{ljk} = +\delta_{il}d!$. Integration over the internal degrees of freedom is the integral $\int d\Omega d^dq/{(2\pi)}^{d+1}$ which is to be computed. The coupling constant $\lambda$ has already been factored out so the factor at A is $i \Omega$, and that at B is $i (\omega-\Omega)$ which we assume are independent of momentum. The internal lines give contributions of the (FE phonon) Green's functions: $A_k G^R(\Omega)$ for the lower and $A_j G^R(\omega-\Omega)$ for the upper parts of the loop respectively.

Evaluating the diagram corresponds to calculating the integral:
\beq
D_{jk} = - A_j A_k \int \frac{d^dq}{{(2\pi)}^d} \frac{d\Omega}{2\pi} G^R(\Omega) G^R(\omega-\Omega) \Omega (\omega-\Omega),
\eeq
with $\chi^{(2)}_{il} = C^2 d! \delta_{il}D_{jk}$. 

The Green's function given by Eq. \eqref{eq:GreensFunctReDefSM} has both real and imaginary parts, so $G^R(\Omega) G^R(\omega-\Omega)$ leads to several terms:
\begin{widetext}
\beqy
G^R(\Omega) G^R(\omega - \Omega) &=& \frac{1}{\omega_q ^2 - \Omega^2 }\frac{1}{\left(\omega_{Q-q}^2 - (\omega-\Omega)^2 \right)} \nonumber \\
&& - \; \frac{\pi^2}{4\omega_q \omega_{Q-q}}\Big[ \delta(\omega_q-\Omega)\delta(\omega_{Q-q}-\omega + \Omega) - \delta(\omega_q + \Omega)\delta(\omega_{Q-q}-\omega + \Omega) \nonumber \\
&& \hskip 2cm - \; \delta(\omega_q - \Omega)\delta(\omega_{Q-q}+\omega -\Omega) + \delta(\omega_q + \Omega)\delta(\omega_{Q-q}+\omega- \Omega) \Big] \nonumber \\
&& + \; \frac{i\pi}{2} \left\{ \frac{\delta(\omega_{Q-q}-\omega + \Omega)  - \delta(\omega_{Q-q} + \omega-\Omega)}{\omega_{Q-q}(\omega_{q}^2 - \Omega^2)} + \frac{\delta(\omega_q-\Omega)  - \delta(\omega_q + \Omega)}{\omega_{q}(\omega_{Q-q}^2 -(\omega-\Omega)^2)} \right\}. \nonumber 
\eeqy
\end{widetext}
The integral over the internal energy is calculated first, then one considers that the momentum dependence of the phonon spectrum is irrelevant near the FE QCP so $\omega_{Q-q} = \omega_q = \omega_0 \sqrt{\delta_x}$. Assuming spherical symmetry, the integral over $d^dq$ becomes $\int_0^\infty dq q^2$ in $d=3$; evaluating up to some cut-off value $\Lambda$ introduces a $\Lambda^3$ weight to $\chi^{(2)}$. 

The real part comes from the middle term of $G^R(\Omega)G^R(\omega-\Omega)$: 
\beq
\mathrm{Re}[\chi^{(2)}] = - \frac{C^2 \delta_{il} A_j A_k\Lambda^3}{8\pi \omega_x} f(\omega)  \label{eq:RealChi2full}
\eeq
with
\beqy
f(\omega) &= & \omega \left[2\delta(\omega) + \delta\left(2\omega_0\sqrt{\delta_x} + \omega\right) - \delta\left(2\omega_0\sqrt{\delta_x} - \omega \right) \right] \nonumber \\
    && + \; \omega_0\sqrt{\delta_x} \left[ \delta\left(2\omega_0\sqrt{\delta_x} - \omega\right) + \delta\left(2\omega_0\sqrt{\delta_x} + \omega\right) \right]. \nonumber \\
\eeqy
The imaginary part, from the first (via Cauchy's residue theorem) and third ($\delta-$functions) terms of $G^R(\Omega)G^R(\omega-\Omega)$, is given in the main text [Eq. \eqref{eq:ImChi2}]. Dimensional analysis gives the dimensions of the $A_k$ factors in $\chi^{(2)}$, and the $n_0^2$ factor in $\chi^{(1)}$ contains an implicit integral over $d\omega d^dq$ to ensure dimensional consistency. For simplicity, $|A_k| = 1$ and $|n_0^2|=1$ are used.

In the limit of static dipoles ($\omega=0$), both the real and imaginary parts of $\chi^{(1)}$ are zero due to the $\omega^2$ factor. Meanwhile, $\mathrm{Re}[\chi^{(2)}]$ diverges with an overall negative factor, thus determining the static behaviour. The imaginary part of $\chi^{(2)}$ also diverges. At the FE QCP where $\delta_x =0$, the real part of $\chi^{(1)}$ is a negative constant and the imaginary part is zero. The real part of $\chi^{(2)}$ goes to $-\infty$ exactly at the FE QCP, but is zero on the approach, and the contribution to $\mathrm{Im}[\chi_m]$ is a positive, energy dependent constant.

In the opposite limit, $\omega \rightarrow \infty$, the real part of $\chi^{(1)}$ is a negative constant, while $\mathrm{Re}[\chi^{(2)}]$ diverges to $-\infty$. The imaginary parts are $\mathrm{Im}[\chi^{(1)}] = 0$ unless $\delta_x = \infty$ too and $\mathrm{Im}[\chi^{(2)}] = 0$ as $1/\omega + 1/\omega^2$. The system is well behaved in that the rate of energy absorption, as quantified by $\mathrm{Im}[\chi_m]$ is finite, even in the limit of infinite energies~\cite{White_QTheorMag}.

\section{IV: $\chi^{(1)}$ and $\chi^{(2)}$ contributions}

\begin{figure}[ht!]
\includegraphics[width=\columnwidth]{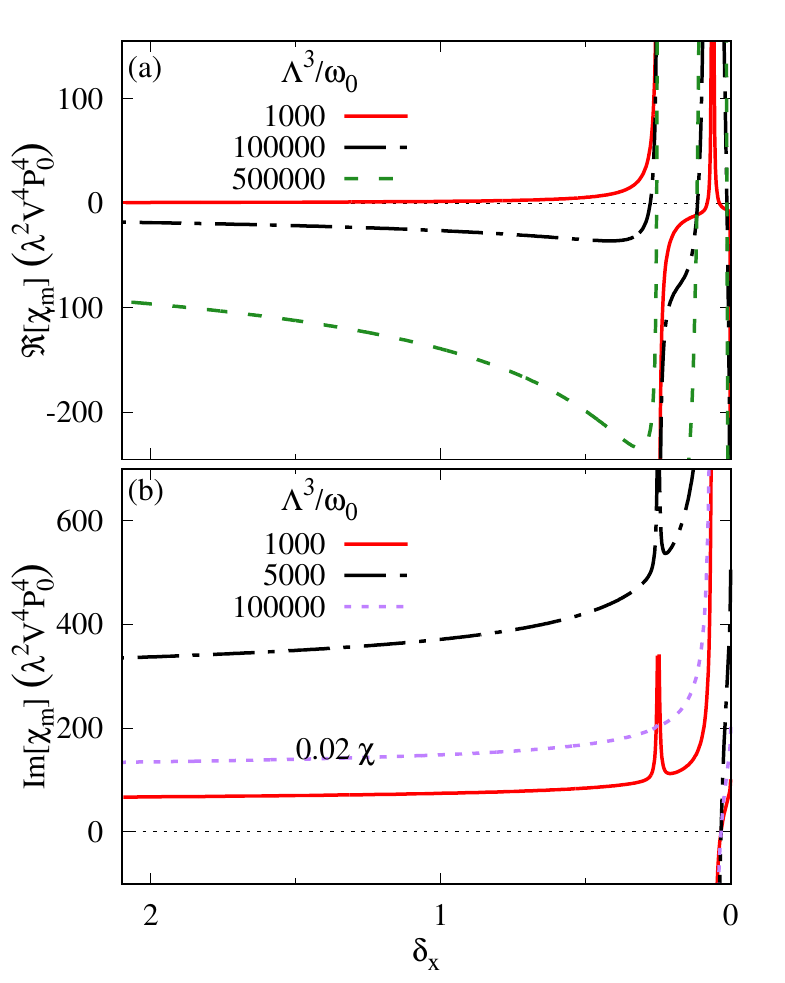}
\caption{(a) Real part of $\chi_m$ at $\omega/\omega_0 =0.5$ for $\Lambda^3/\omega_0 = 1\times 10^3,\; 1 \times 10^5, \; 5\times 10^5$. Reducing the significance of the $\chi^{(2)}$ contribution narrows the width of the peak in $\mathrm{Re}[\chi_m]$ near the FE QCP and ensures diamagetic behaviour ($\mathrm{Re}[\chi_m]>0$ but small) at larger distances from the QCP. (b) Imaginary part of $\chi_m$ at $\omega/\omega_0 = 0.5$ for several values of $\Lambda^3/\omega_0 = 1\times 10^3,\; 5 \times 10^3, \; 1\times 10^5$; for the last $2\times 10^{-2} \mathrm{Im}[\chi_m]$ is plotted: the position of the peak at $\delta_x = \sqrt{\omega/\omega_0}$ does not move, but it becomes less noticeable as $\Lambda^3/\omega_0$ is increased. \label{fig:varChi2Weight}} 
\end{figure}

The contribution from $\chi^{(2)}$ depends on a momentum-dependent factor $\Lambda^3/\omega_0$. As seen in Fig. \ref{fig:varChi2Weight}, the regions of positive $\chi_m$ after the initial divergence at $\delta_x = \omega^2/\omega^2_0$ are suppressed for small $\Lambda^3/\omega_0$,  corresponding to large distances (or sample size) for a given phonon frequency $\omega_0$. The observation of these features close to the FE QCP will depend strongly on the distances and energies considered.

\begin{center}
\begin{figure}[ht!]
\includegraphics[width=0.9\columnwidth]{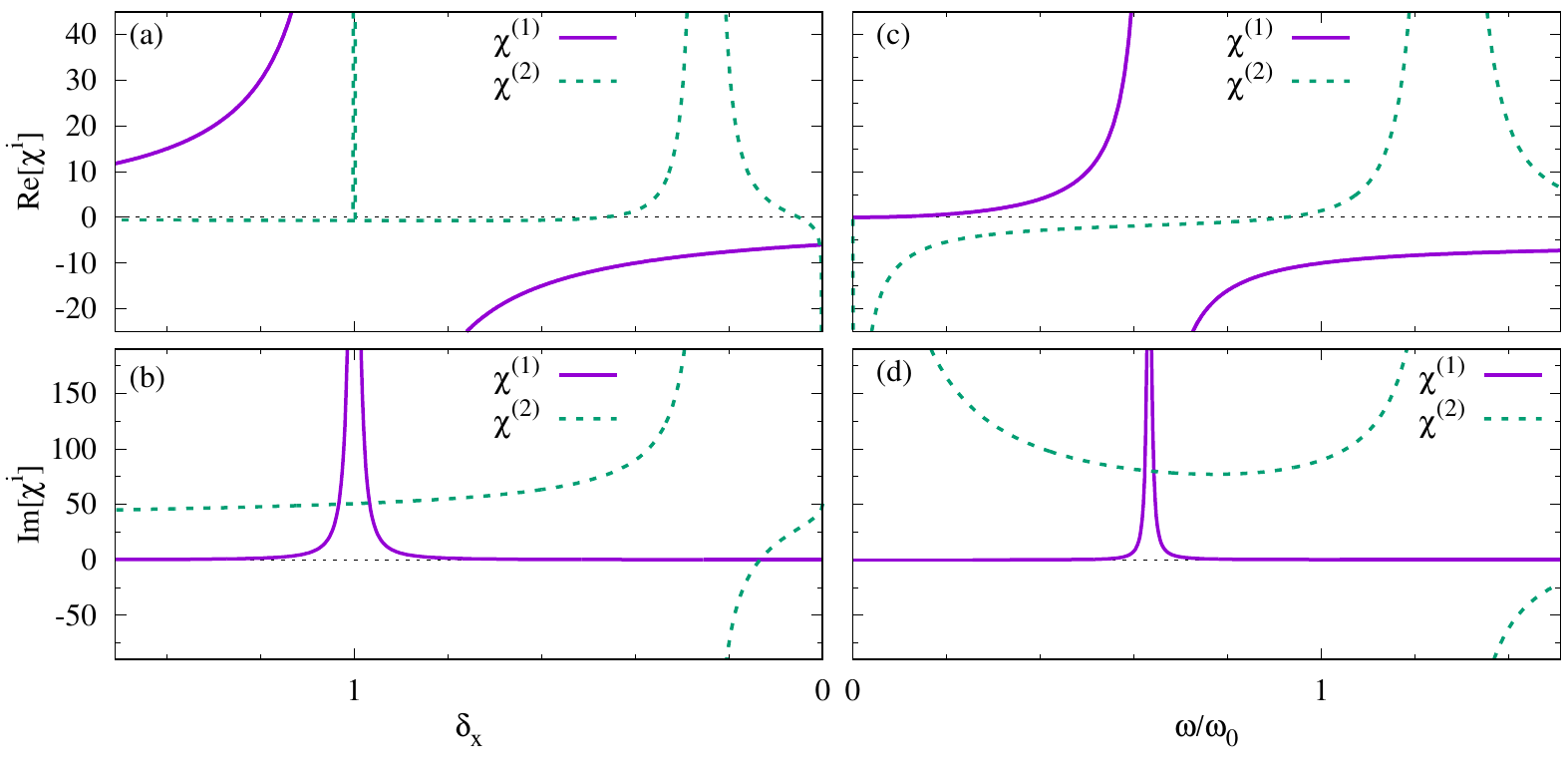}
\caption{Real and imaginary parts of $\chi^{(1)}$ (solid purple lines) and $\chi^{(2)}$ (dashed green lines) contributions to $\chi_m$. (a), (b) as a function of distance from the FE QCP ($\delta_x=0$) at fixed $\omega= \omega_0$; scale of $\chi^{(2)}$ is $5000$. (c), (d) as a function of energy at fixed $\delta_x = 0.4$; scale of $\chi^{(2)}$ is $1000$, see Fig. \ref{fig:varChi2Weight} for the effect of changing weight of $\chi^{(2)}$ on the total susceptibility. In all cases, the scale is in terms of the common size $\lambda^2 V^4 P_0^4$, and the finite width and height of the peaks in $\mathrm{Re}[\chi^{(2)}]$ and $\mathrm{Im}[\chi^{(1)}]$ are the result of replacing the $\delta-$functions with Lorentzian functions.
\label{fig:SeparateChiContribsAll4}}
\end{figure}
\end{center}

The contributions of $\chi^{(1)}$, given by the Green's function, Eq. \eqref{eq:GreensFunctReDefSM}, and $\chi^{(2)}$, Eq. \eqref{eq:RealChi2full} here and Eq. \eqref{eq:ImChi2} of the main text, are plotted as functions of $\delta_x$ in Fig. \ref{fig:SeparateChiContribsAll4} (a) and (b), and as functions of energy in Fig. \ref{fig:SeparateChiContribsAll4} (c) and (d). This highlights the origin of the features seen in Figs. 2 and 3 of the main text. In all plots, both here and the main text, the $\delta$-functions from $\mathrm{Im}[G^R(\omega)]$ have been replaced by Lorentzian functions.

\section{V: Behaviour in a magnetic field}

The additional term in the Lagrangian describing the interaction between a magnetic moment and an external field, $L_{B} = \mathbf{B}\cdot \mathbf{m} = \lambda \mathbf{B}\cdot(\mathbf{p}\times \partial_t \mathbf{p})$, can be treated as a perturbative term in the full Lagrangian, the second order expansion of which gives the diagram
\begin{center}
\includegraphics[width=0.67\columnwidth]{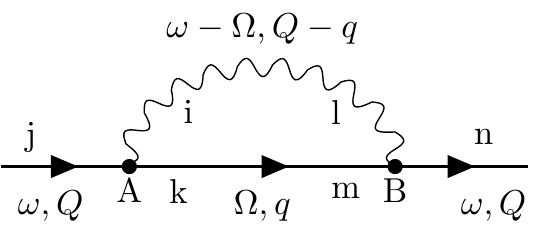}\\
\end{center}
Using the standard diagrammatic rules, this corresponds to the integral:
\beq
I_B = L \omega \int \frac{d^dq}{(2\pi)^d}\frac{d\Omega}{2\pi} \Omega G^R(\Omega,q)G_B(\omega-\Omega,Q-q),
\eeq
where the overall positive sign is from $(i^2) \epsilon_{ijk}\epsilon_{ikn}$, and $L \equiv C^2 A_k A_n d! \delta_{jn}$. $G^R$ is the usual Green's function for the ferroelectric propagator, Eq. \eqref{eq:GreensFunctReDefSM}, and $G_B$ is the magnetic propagator:
\beqy
G_B(\omega-\omega^\prime, q-q^\prime) &\equiv& \langle B_i (-\omega, -q) | B_l(\omega^\prime, q^\prime)\rangle \nonumber \\
& =& B^2 \delta_{il} \delta(\omega-\omega^\prime) \delta(q-q^\prime).
\eeqy
In the diagram, have $\omega-\Omega$, and $Q-q$. Thus the calculation to be performed is:
\beqy
I_B^{jn} &=& L \omega B^2 \int \frac{d^dq}{{(2\pi)}^d} \frac{d\Omega}{2\pi} \delta(\omega - \Omega) \delta(Q-q) \Omega \times \nonumber \\
&& \times\; \left( \frac{1}{\omega_q^2 - \Omega^2} + \frac{i\pi}{2\omega_q}\left[ \delta(\omega_q-\Omega)  - \delta(\omega_q + \Omega) \right]\right) \qquad
\eeqy
The $\delta$-functions from $G_B$ lead to evaluation of the integrals at $\Omega = \omega$ and $q = Q$. The integral over energy is straightforward, and after assuming spherical symmetry, $d^dq = dq q^{d-1} S_d$; with $S_3 = 4\pi$, one finds
\beqy
I_B^{jn} 
&=& \frac{L \omega^2 Q^2 B^2}{4\pi^3} G^R(\omega, Q).
\eeqy
A magnetic field therefore affects the energy of the FE phonons, but does not, at this level of approximation, move the FE QCP.

\onecolumngrid
\twocolumngrid

\end{document}